# Intra-bunch feedback system developments at DAFNE*

A. Drago, INFN-LNF, Frascati and Tor Vergata University, Rome, Italy

D. Alesini, S. Caschera, A. Gallo, INFN-LNF, Frascati, Italy

J.D. Fox, Stanford University, Stanford, USA

J. Cesaratto (Toohig Fellow), J. Dusatko, J. Olsen, C. Rivetta, O. Turgut, SLAC, Menlo Park, USA

W. Hofle, G. Iadarola, K. Li, E. Metral, E. Montesinos, G. Rumolo, CERN, Geneva, Switzerland

S. De Santis, M. Furman, J-L Vay, LBNL, Berkeley, USA

M. Tobiyama, KEK, Tsukuba, Japan

*Abstract*

This paper presents history and evolution of the intra-bunch feedback system for circular accelerators. This project has been presented by John D. Fox (SLAC/Stanford Un.) at the IPAC2010 held in Kyoto. The idea of the proposal is to build a flexible and powerful instrument to mitigate the parasitic e-cloud effects on the proton (and potentially positron) beams in storage rings. Being a new and ambitious project, the financial issues have been quite important. US LHC Accelerator Research Program (LARP) and other institution funding sources have assured the development of the design for implementing the feedback in the SPS ring at CERN. Here the intra-bunch feedback system has been installed and tested in the frame of the LIU (LHC Injector Upgrade) program.

After the end of the LARP funding, a possible new interesting chance to continue the R&D activity, could be by implementing the system in a lepton storage ring affected by e-cloud effects. For achieving this goal, a possible experiment could be carried out in the positron ring of DAFNE at Frascati, Italy. The feasibility of the proposal is evaluated in the following sections. In case of approval of the experiment, indeed the project could be inserted in the DAFNE-TF (DAFNE Test Facility) program that is foreseen after the 2020 for the following 3-5 years.

## INTRODUCTION

As it is well known, in a storage ring the photons emitted by the beams and hitting the vacuum chamber form electron clouds. These negatively charged clouds affect the beam dynamics of the positively charged beams with several undesired effects.

Many mitigation techniques are studied and developed. One of these is the "intra-bunch feedback", that has been often called "e-cloud feedback" or "wide band feedback system" (WBFS), too.

The intra-bunch feedback definition derives from the fact that it is a kind of system that treats the bunch of particles not as a rigid body as in the bunch-by-bunch feedback but considering each bunch split in several slices. Hence, the system applies an individual correction signal kick to each bunch slice. It is noteworthy to underline that a standard bunch-by-bunch system working on the centroid of the bunch is still usually necessary.

Quite the opposite, the wideband name has been chosen for the fundamental importance to have a wide frequency band for this kind of feedback. By using this definition, the main technological feature and challenge is reported in the name.

The e-cloud feedback definition just remembers the main goal of the system, i.e. to mitigate the e-cloud effects on the beams.

## A LARGE COLLABORATION

After the year 2009, John Fox started to ask funds to the LARP program in USA for research on a new type of feedback, able to damp bunch slices independently and to be implemented in SPS/LHC. The LARP (U.S. LHC Accelerator Research Program) consists of four laboratories, Brookhaven National Laboratory (BNL), Fermi National Accelerator Laboratory (FNAL), Lawrence Berkeley National Laboratory (LBNL) and Stanford Linear Accelerator Center (SLAC).

They collaborate with CERN and other institutions in the context of the High Luminosity LHC program (HL_LHC) on the Large Hadron Collider in order to:
a) Make more luminosity, in an early stage.
b) Collaborate in interaction region upgrades, to make more luminosity, in a subsequent stage.
c) Use, develop, and preserve unique U.S. resources and capabilities.

Unfortunately, the LARP funding to the WBFS ended on September 2017.

The first general talk on the project was presented at the IPAC'10 conference held at Kyoto and John Fox proposed an "SPS E-cloud Feedback" [1]. Along the following years, a large collaboration grew with the interest of experts from many laboratories around the world, as reported in the Annex A.

The first goal of the project was to implement the new feedback in SPS ring at CERN as part of the LIU (LHC

Injector Upgrade) program. A prototype of the WBFS was developed and tested on the SPS proton beam and the proof of principle was achieved with single bunch as shown in Figure 1, and reported in [2].

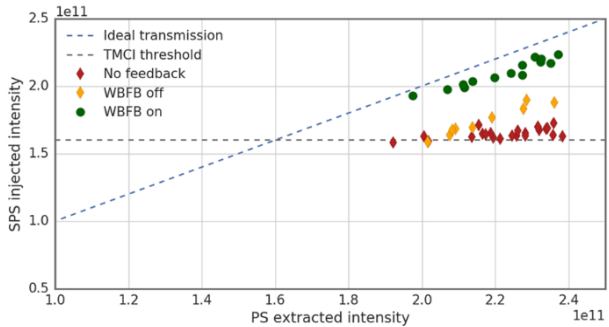

Figure 1: The effect of the WBFS on SPS beam current, from K. Li talk at LIU MD Days, 15/3/2018 [2].

The main R&D activities are outlined in the following points:
a) to characterize the pickups installed in the SPS ring in the past to figure out if they showed the needed frequency band and flatness for the bunch signals.
b) To develop a new feedback digital unit able to process 8 or 4 Gsamples/s [3-4] and to output the computed correction signal for each slice, see Figure 2.
c) To identify power amplifiers with a bandwidth of 1 GHz and showing a correct 2 ns impulse response.
d) To carry on R&D efforts to evaluate the best wide band kicker (1 GHz) to which applying the correction signal.
e) To install and test the new WBFS on the SPS proton beam.

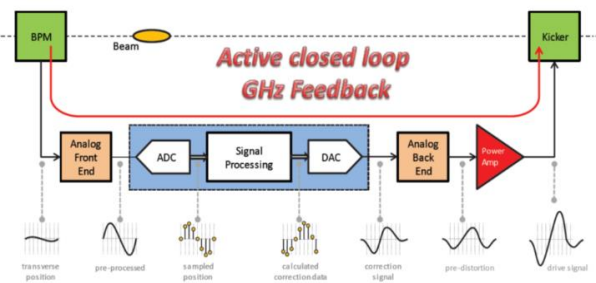

Figure 2: Layout of the WBFS, from [3].

The main efforts of the collaboration were shared between the interested laboratories depending on the skill of the researchers. However, the b) and d) activities have requested the more important R&D manpower. The b) task, to develop the digital processing unit, was carried on mainly by the SLAC team, while the d) activity, kicker studies, were shared between LBNL, CERN, SLAC and INFN/LNF teams. Regarding the c) task (power amplifier), the Japanese contribution was relevant to identify the best solution. Of course the e) task, installation and test at SPS, required big efforts from the CERN feedback team.

It is important to highlight for the following discussion, that in the SPS storage ring the intra-bunch feedback processes proton bunches having a length of 1.7 ns and divides the bunch in 16 slices. These features are reported in the O. Turgut's talk and paper for the IPAC16 [5-6].

## DAFNE TEAM CONTRIBUTIONS

At LNF, the Italian laboratory, from the end of 2011, some researchers and engineers of the DAFNE team start to collaborate with the SLAC/CERN/LBNL task force as participating to Hi-Lumi-LHC collaboration funded by EU (November 2011-2015, within the FP7, seventh Framework Programme) to increase the LHC luminosity.

In particular efforts were carried on to evaluate the more suitable kicker to be implemented considering the large frequency bandwidth of the signals. Several designs were considered [7].

In particular, three different kicker types were studied and compared:
a) 10 cm long stripline kicker designed to have a large frequency bandwidth by S. De Santis at ALS Berkeley (installed in SPS in 2016);
b) the slotted kicker studied at the beginning mainly by J. Cesaratto;
c) an approach based on several RF cavities proposed by A. Gallo from LNF.

The LNF team was involved only in the b) and c) cases.
Indeed, John M. Cesaratto spent few weeks at LNF in 2013 working on the slotted kicker and funded by the FAI ("Fondo Affari Internazionali") program of INFN. The work done analysed three different designs versus performance and impedance of the kicker [8], see Figure 3.

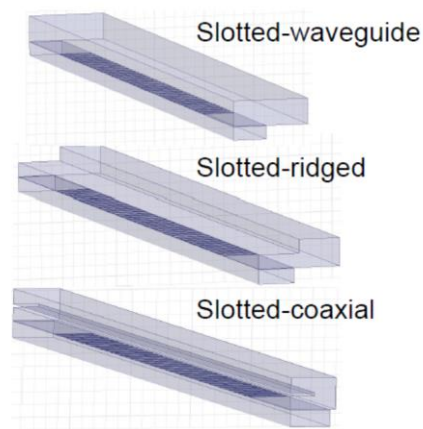

Figure 3: Quarter models of three slotted kickers studied.

The slotted-type kicker geometries, evaluated by HFSS, are similar to that used for the stochastic cooling:
a) the slotted-waveguide kicker consists of a waveguide coupled to a beam pipe via slots.

b) The slotted-ridged waveguide is an extension of the slotted-waveguide kicker where the ridge concentrates the field.

c) The slotted-coaxial kicker has a coaxial transmission line within the waveguide.

In the Figure 3 the quarter model geometries of three slotted kickers are presented.

The study brought eventually to an engineered version of the c) option, having 1 m of length, later designed at CERN to be installed in SPS ring in the June 2018. In this case, the slotted kicker frequency band is complementary to the one of the stripline kickers, already installed in the past in SPS. Indeed, the slotted kicker achieves a better response at the higher frequency, between 0.7 and 1 GHz.

Going to the c) option, the RF cavities based, it has remained at a preliminary stage up to now. This approach proposed by A. Gallo [9-10] wishes to operate on the bunch by splitting the correction signal in several frequency bands that need to be applied at RF cavities and stripline kickers.

Even if it is a very interesting approach, this design was not carried on up to now most likely for the difficulty to equalize and to time the different frequency bands of the bunch correction signal as well as for the impedance added to the ring. Nevertheless, it could be developed and installed in the next years in SPS if funded by a new R&D program.

As shown in the Figure 4, the correction signal, after the equalization and the Fourier transform expansion, is split and applied respectively to a cavity or stripline kicker working up to f0 (400 MHz) and to other two cavities working around 2f0 (800 MHz) and 3f0 (1200 MHz), both using the TM110 deflecting mode.

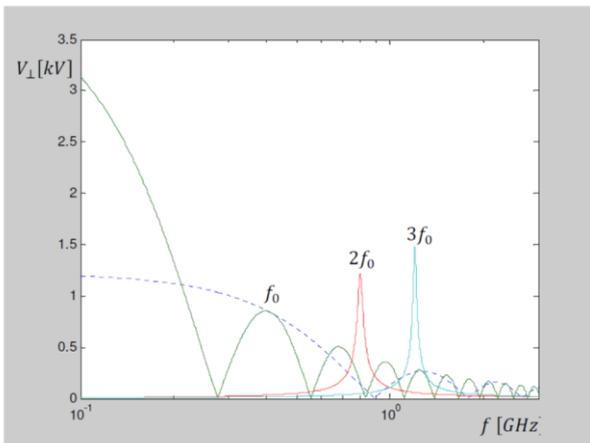

Figure 4: Deflecting voltages delivered by striplines and TM010 single cell cavities excited with 1 kW RF power.

## A FEASIBILITY STUDY FOR DAFNE-TF

Even if at the present no funds from SLAC and LNF are available for new activities, at LNF in the December 2018, an ICFA and ARIES Mini-Workshop on DAFNE as Open Accelerator Test Facility (DAFNE-TF) will be held [11].

The workshop is intended to discuss the interest from scientists to access the DAFNE e+ e- complex, which will conclude its physics program as collider in 2020. An infrastructure almost unique, that could open as test facility to the international community for studies of accelerator technologies and beam physics, for small experiments, and to be used as a test bed for enterprises active in the sector of components for accelerators.

Given that the intra-bunch feedback up to now has never been implemented for lepton storage rings, a proposal for DAFNE-TF can be drawn up. Of course, in case of approval, it needs to be funded.

The interest in a lepton storage ring would be not only in the e+ ring to cope the e-cloud effects but also in the e- ring for a possible application to mitigate the Transverse Mode Coupling Instability (TMCI).

The main limitation for using an intra-bunch feedback in DAFNE is in the too short length of the bunch. With the present configuration, this is about 300 ps. Given that the main RF frequency is ~368 MHz, each bucket is ~2.7 ns, with a semi-period of 1.35 ns. From a comparison with the SPS bunch length seen above, it is necessary to stretch the DAFNE bunch as more as possible.

With respect to SPS beam using the intra-bunch feedback in the vertical plane, at DAFNE the e-cloud effects require a system in the horizontal plane where the instability is stronger than in the vertical plane [12-13]. The intra-bunch feedback could help to store more beam current than the 1.2A that has been achieved for the e+ ring in DAFNE up to now.

A beam dynamics simulation study needs to be carried on in order to be able to store in the ring a very long bunch (~1.0-1.3 ns). This value for the bunch length seems feasible, most likely by using a third harmonic cavity to stretch the bunches. An issue could arise to find space for the new cavity. It most likely should be placed in the second interaction region.

Going to the feedback design, it has to be added to the other three systems (horizontal, vertical and longitudinal). At the present in the DAFNE e+ ring there is a second horizontal feedback with a stripline kicker. This complementary system could be taken off and a slotted kicker can be placed where now is located the second horizontal stripline kicker. The available space is about one meter.

A 4 GSamples/s processing unit is necessary and maybe it can be borrowed from SLAC or from CERN.

Otherwise it needs to be designed again partially or fully. At LNF there are other data acquisition systems that can be adapted reprogramming the FPGA code and making some small hardware modifications [14]. See in Figure 5 the eight data acquisition systems working in parallel for the experiment 3+L in DAFNE e+ ring.

As third option, a completely new digital system should be designed. This solution would have the advantage to use state-of-art components, even if it will be more expensive.

As starting point, a number of 8 slices (8 channels for each bunch) seems to be sufficient as processing power. However, the SPS system implements the double, 16 channels, as reported above.

Furthermore, for budget reasons, it is necessary to foresee the necessary power. The power amplifiers are the most expensive items in the feedback. At the present, in the DAFNE e+ ring the bunch-by-bunch systems have last stage amplifiers of 2x250W in the horizontal plane, and the same in the vertical one. A second horizontal bunch-by-bunch feedback is ready but still to be tested with the beam, being equipped with 2x500W amplifiers. For the intra-bunch, 250 W power seems reasonable to begin the experiment, even if it could be useful to foresee a later increment of the power up to 1kW.

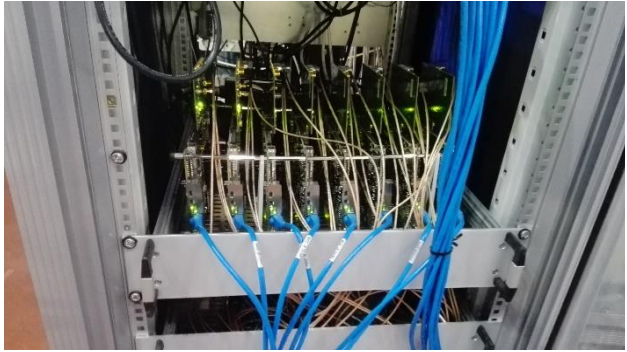

Figure 5: DAFNE 3+L data acquisition system based on 8 FPGA based boards working in parallel.

After all the previous considerations, the funds to implement the WBFS in DAFNE e+ ring would be of the order of 300k euro, without considering the costs for the manpower.

## CONCLUSION

Transverse instabilities produced by e-cloud effect can limit both beam and single bunch currents in proton and positron beams. The wideband feedback has already been demonstrated at SPS that it is a powerful instrument to mitigate the intra-bunch instabilities.

Nevertheless, there is no more funding from LARP and Hi-Lumi-LHC to continue the R&D activities.

After the year 2020, DAFNE-TF could offer the occasion to implement a wideband feedback system in a positron ring. Given the beam current limits in DAFNE for the e-cloud effects in the horizontal plane, it would be more interesting to design it for this plane.

As consequence, the kicker design should be different from SPS where it is working in the vertical plane.

At a first attempt, the budget for the experiment could be limited to about 300k€. In case of approval of the project, studies for understanding how to stretch the bunches should be carried on as soon as possible to have a correct evaluation of the beam characteristics, remembering that a too short bunch cannot be useful for the intra-bunch feedback system.

## ACKNOWLEDGEMENT

This work was supported in part by the European Commission under the HORIZON2020 Integrating Activity project ARIES, grant agreement 730871.

# ANNEX A:

## THE WBFS COLLABORATION*

J.D. Fox, J. Cesaratto (Toohig fellow), J. Dusatko, J. Goldfield, J. Olsen, M. Pivi, K. Pollock, N.Redmon, C. Rivetta, O. Turgut, S. Uemura, SLAC, 2575 Sand Hill Rd, Menlo Park, CA 94025, USA

D. Aguilera, G. Arduini, H. Bartosik, S. Calvo, W. Hofle, G. Iadarola, G. Kotzian, K. Li, E. Metral, E. Montesinos, G. Rumolo, B. Salvant, U. Wehrle, M. Wendt, C. Zanini, CERN, 1211 Geneva 23, Switzerland

S. De Santis, M. Furman, H. Quian, M. Venturini, J-L Vay, LBNL, Lawrence Berkeley National Laboratory, 1 Cyclotron Rd, Berkeley, CA 94720, USA

D. Alesini, A. Drago, A. Gallo, F. Marcellini, M. Zobov, INFN-LNF, Via Enrico Fermi 40, 00044 Frascati, Italy

M. Tobiyama, KEK, Tsukuba, Ibaraki 305-0801, Japan